\documentclass[%
reprint,
superscriptaddress,
showpacs,preprintnumbers,
 amsmath,amssymb,
 aps,
]{revtex4-1}

\usepackage{comment}
\usepackage{graphicx}
\usepackage{dcolumn}
\usepackage{bm}
\usepackage{enumitem}
\usepackage{color}
\usepackage{multirow}

\newcommand{\fig}[1]{Fig.~\ref{#1}}
\newcommand{\tab}[1]{Table~\ref{#1}}
\newcommand{\Sec}[1]{Sec.~\ref{#1}}
\newcommand{\eq}[1]{Eq.~(\ref{#1})}
\newcommand{\Ref}[1]{Ref.~\cite{#1}}

\begin{document}

\preprint{
}
\title{Determination of $\gamma$-ray widths in $^{15}$N using nuclear resonance fluorescence}

\author{T.~Sz\"ucs}%
\email{t.szuecs@hzdr.de}%
\affiliation{Helmholtz-Zentrum Dresden - Rossendorf (HZDR), D-01328 Dresden, Germany}

\author{D.~Bemmerer}%
\affiliation{Helmholtz-Zentrum Dresden - Rossendorf (HZDR), D-01328 Dresden, Germany}

\author{A.~Caciolli}%
\affiliation{Dipartimento di Fisica e Astronomia, Universita di Padova, I-35131 Padova, Italy}
\affiliation{INFN-Laboratori Nazionali di Legnaro, I-35020 Legnaro, Italy}

\author{Zs.~F\"ul\"op}%
\affiliation{Institute for Nuclear Research (MTA Atomki), H-4001 Debrecen, Hungary}

\author{R.~Massarczyk}%
\altaffiliation[Present address: ]{Los Alamos National Laboratory, Physics Division, Los Alamos, New Mexico 87545, USA}%
\affiliation{Helmholtz-Zentrum Dresden - Rossendorf (HZDR), D-01328 Dresden, Germany}
\affiliation{Technische Universit\"at Dresden, D-01069 Dresden, Germany}

\author{C.~Michelagnoli}%
\affiliation{Grand Acc\'el\'erateur National d'Ions Lourds (GANIL), F-14076 Caen, France}

\author{T.~P.~Reinhardt}%
\affiliation{Technische Universit\"at Dresden, D-01069 Dresden, Germany}

\author{R.~Schwengner}%
\affiliation{Helmholtz-Zentrum Dresden - Rossendorf (HZDR), D-01328 Dresden, Germany}

\author{M.~P.~Tak\'acs}%
\affiliation{Helmholtz-Zentrum Dresden - Rossendorf (HZDR), D-01328 Dresden, Germany}
\affiliation{Technische Universit\"at Dresden, D-01069 Dresden, Germany}

\author{C.~A.~Ur}%
\affiliation{Extreme Light Infrastructure - Nuclear Physics (ELI-NP), R-077125 Magurele, Romania}

\author{A.~Wagner}%
\affiliation{Helmholtz-Zentrum Dresden - Rossendorf (HZDR), D-01328 Dresden, Germany}

\author{L.~Wagner}%
\affiliation{Helmholtz-Zentrum Dresden - Rossendorf (HZDR), D-01328 Dresden, Germany}
\affiliation{Technische Universit\"at Dresden, D-01069 Dresden, Germany}

\date{Published 21 July 2015. Addendum published xx October 2015}

\begin{abstract}
\begin{description}
\item[Background] The stable nucleus $^{15}$N is the mirror of $^{15}$O, the bottleneck in the hydrogen burning CNO cycle. Most of the $^{15}$N level widths below the proton emission threshold are known from just one nuclear resonance fluorescence (NRF) measurement, with limited precision in some cases. A recent experiment with the AGATA demonstrator array determined level lifetimes using the Doppler shift attenuation method in $^{15}$O. As a reference and for testing the method, level lifetimes in $^{15}$N have also been determined in the same experiment.
\item[Purpose] The latest compilation of $^{15}$N level properties dates back to 1991. The limited precision in some cases in the compilation calls for a new measurement to enable a comparison to the AGATA demonstrator data. The widths of several $^{15}$N levels have been studied with the NRF method. 
\item[Method] The solid nitrogen compounds enriched in $^{15}$N have been irradiated with bremsstrahlung. The $\gamma$-rays following the deexcitation of the excited nuclear levels were detected with four  high-purity germanium detectors.
\item[Results] Integrated photon-scattering cross sections of 10 levels below the proton emission threshold have been measured. Partial gamma-ray widths of ground-state transitions were deduced and compared to the literature. The photon scattering cross sections of two levels above the proton emission threshold, but still below other particle emission energies have also been measured, and proton resonance strengths and proton widths were deduced.
\item[Conclusions] Gamma and proton widths consistent with the literature values were obtained, but with greatly improved precision.
\end{description}
\end{abstract}


\maketitle

\section{\label{sec:intro} Introduction}

The $^{15}$N nucleus is the mirror nucleus \cite{Hodgson97-book} of $^{15}$O, the reaction product of the slowest reaction, $^{14}$N(p,$\gamma$)$^{15}$O, in the hydrogen burning CNO cycle \cite{Iliadis07-book}. The reaction rate of this reaction has a major influence on the determined age of some very old globular clusters \cite{Imbriani04-AA}, and it also has a key importance in the prediction of the solar CNO neutrino flux with the standard solar model (SSM) \cite{Haxton08-AJ}.
In particular, the gamma width of the $E_x = 6.792$\,MeV, $3/2^+$ level in $^{15}$O strongly affects the rate of the CNO cycle \cite{Adelberger11-RMP}. This level is the isospin mirror of the $E_x = 7.301$\,MeV, $3/2^+$ level in $^{15}$N \cite{Adelberger98-RMP}.

A recent experiment with the advanced gamma tracking array (AGATA) demonstrator aimed to measure level lifetimes in $^{15}$O \cite{Michelagnoli13-Diss,Michelagnoli-prep} using the Doppler shift attenuation method (DSAM). 
Owing to the very good angular resolution of AGATA when compared to standard single crystal  high-purity germanium (HPGe) detectors, and to the high recoil velocity because of the inverse kinematics, the line shapes of the Doppler shifted peaks become examinable at several angles, the lifetimes of the levels were determined from the best fits. Transitions in $^{15}$N have also been analyzed in the same experiment for testing the method \cite{Michelagnoli13-Diss,Michelagnoli-prep}. In some cases the $^{15}$N level widths in the latest compilation from 1991 \cite{AjzenbergSelove91-NPA} are not known with a sufficient precision to be a clear reference.

Most of the level widths in the compilation \cite{AjzenbergSelove91-NPA} are based on just one nuclear resonance fluorescence (NRF) measurement from 1981 \cite{Moreh81-PRC}. 
The aim of the present work is to improve the precision of the level widths by using an efficient low-background photon scattering setup \cite{Schwengner05-NIMA}, and thus providing better reference data for the DSAM measurement.

The paper is organized as follows: Experimental details are described in Secs. \ref{sec:method} and \ref{sec:measurement}; the resulting gamma widths are presented in \Sec{sec:results}, where also a comparison to literature data is made; \Sec{sec:sum} provides the conclusions and a short summary.

\section{\label{sec:method} Experimental technique and setup used}

\begin{figure}[t]
\includegraphics[width=0.99\columnwidth]{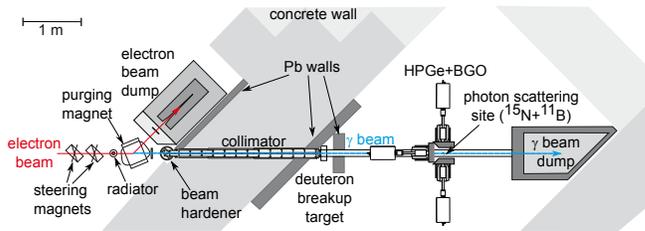}
\caption{\label{fig:setup} (Color online) Schematic top view of the photon scattering facility \cite{Schwengner05-NIMA} at the ELBE superconducting electron accelerator of HZDR. Detectors at 127$^\circ$ with respect to the $\gamma$-beam are below and above the beam line.}
\end{figure}

The widths of several levels in $^{15}$N were studied at the photon scattering facility \cite{Schwengner05-NIMA} at the superconducting electron accelerator ELBE \cite{Gabriel00-NIMB,Teichert03-NIMA} of Helmholtz-Zentrum Dresden - Rossendorf (HZDR). A schematic view of the whole setup is presented in \fig{fig:setup}.
Bremsstrahlung was produced with an electron beam impinging onto a niobium radiator foil. The photons scattered by the $^{15}$N NRF target were detected by four HPGe detectors.

\subsection{\label{sec:photon} The nuclear resonance fluorescence technique}

In case of an NRF experiment the $\gamma$-ray rate $R_\gamma(E_\gamma,\theta)$ observed at an angle $\theta$ with respect to the beam is proportional to the energy- and solid-angle integrated resonant scattering cross section ($I_\sigma$) of the excited state $x$:
\begin{equation}
I_\sigma(0\rightarrow E_x \rightarrow E_f) = \frac{R_\gamma(E_\gamma,\theta)}{\epsilon(E_\gamma) W(E_\gamma,\theta) \Phi(E_x) N_N} ,
 \label{eq:I_from_detection}
\end{equation}
where $E_x$ and $E_f$ are the energies of the excited and final states, $\epsilon(E_\gamma)$ is the absolute detection efficiency of the $\gamma$ ray with given energy, $W(E_\gamma,\theta)$ is the angular correlation of the excitation and the de-excitation transitions, $\Phi(E_x)$ is the photon flux at the energy of the level, and $N_N$ is the number of target nuclei per unit area.

On the other hand the integrated resonant scattering cross section is related to the gamma width by the following relation:
\begin{equation}
I_\sigma(0\rightarrow E_x \rightarrow E_f) = \frac{2J_x+1}{2J_0+1} \left( \frac{\pi\hbar c}{E_x} \right)^2 \frac{\Gamma_0\Gamma_f}{\Gamma_\gamma} ,
 \label{eq:I_from_width}
\end{equation}
where $J_0$ and $J_x$ are the spins of the ground and the excited state, respectively. $\Gamma_0$ and $\Gamma_f$ denote the partial gamma widths of the excited level to the ground and to the final level, respectively, while $\Gamma_\gamma$ is the total gamma width of the level, i.\,e., equal to the sum of all partial gamma widths.
Using the branching ratio $b_f = \Gamma_f / \Gamma_\gamma$  the last factor in \eq{eq:I_from_width} can be expressed as 
\begin{equation}
\frac{\Gamma_0\Gamma_f}{\Gamma_\gamma} = b_f \Gamma_0 .
 \label{eq:I_branc}
\end{equation}

In case of elastic photon scattering, the final state is the ground state, therefore \eq{eq:I_from_width} becomes
\begin{equation}
I_\sigma(0\rightarrow E_x \rightarrow 0) = \frac{2J_x+1}{2J_0+1} \left( \frac{\pi\hbar c}{E_x} \right)^2 b_0 \Gamma_0 .
 \label{eq:I_elastic}
\end{equation}
Because the branching ratios for light nuclei are often available in the literature, the ground-state gamma width can be deduced from the scattering cross sections.

In the following the determination of the different factors in the denominator of \eq{eq:I_from_detection} will be described.

\subsection{\label{sec:detector} $\gamma$-ray detection setup}

All four HPGe detectors used in this study have 100\,\% relative efficiency. They are equipped with bismuth germanate (BGO) scintillators, as active shielding to reduce the continuum background. Lead collimators 10\,cm thick and 3-cm thick lead side shields were applied at each detector to reduce the environmental and photon beam induced background. Two detectors were placed at 127$^\circ$ and two at 90$^\circ$ with respect to the incident $\gamma$ beam. They were located at a distance of 32\,cm and 28\,cm to the target, respectively. 
The flux of low-energy photons entering the detector collimators was suppressed by 3-mm (8\,mm) thick lead and 3-mm thick copper absorbers for the detectors at 127$^\circ$ (90$^\circ$), respectively.

The absolute detection efficiencies of the HPGe detectors were measured up to 2.45\,MeV using calibrated radioactive sources. The obtained absolute efficiencies for the two detectors at each given angle were compatible within the error bars. In the analysis the spectra recorded from two detectors at a given angle were added.

The whole setup including the HPGe detectors, the BGO shields, the collimators, the lead shieldings, the target, and detector holders was implemented in a GEANT4 \cite{Agostinelli03-NIMA} simulation. The reliability of the simulation was previously tested by comparing simulated spectra with measured ones \cite{Rusev08-PRC,Schwengner07-PRC,Massarczyk12-PRC}.

The shape of the measured efficiency curves were consistent with the simulated ones, and the absolute scale agreed within 2\%. Later in the analysis the efficiency curve calculated with GEANT4 and scaled to the absolute experimental values was used. An efficiency uncertainty of 2\% was adopted.

\subsection{\label{sec:ang} Angular correlation}
\begin{table*}[t]
\caption{\label{tab:angular_corr_15N} Angular correlations for $^{15}$N transitions used in the analysis. Level parameters and mixing ratios from \Ref{AjzenbergSelove91-NPA}.} 
\begin{ruledtabular}
\begin{tabular}{*{7}{c}}
$E_x$ (MeV)		& $J^\pi_x$	& $E_f$ (MeV)		& $J^\pi_f$	&Mixing ratio\footnotemark[1]& $W(127^\circ)$	& $W(90^\circ)$		\\
\colrule
\phantom{1}5.298	& $1/2^+$	& 0.000	& $1/2^-$	&						& 1.000			& 1.000	\\
\phantom{1}6.324	& $3/2^-$	& 0.000	& $1/2^-$	& -0.132\,$\pm$\,0.004			& 1.021\,$\pm$\,0.001	& 0.750\,$\pm$\,0.004	\\
\phantom{1}7.301	& $3/2^+$	& 0.000	& $1/2^-$	& +0.017\,$^+_-\,^{0.008}_{0.005}$	& 1.009\,$\pm$\,0.001	& 0.890\,$^+_-\,^{0.006}_{0.004}$	\\
\phantom{1}8.313	& $1/2^+$	& 0.000	& $1/2^-$	& 						& 1.000			& 1.000	\\
\phantom{1}8.571	& $3/2^+$	& 0.000	& $1/2^-$	& +0.085\,$^+_-\,^{0.009}_{0.005}$	& 1.005\,$\pm$\,0.000	& 0.940\,$^+_-\,^{0.006}_{0.003}$	\\
\phantom{1}8.571	& $3/2^+$	& 5.270	& $5/2^+$	& +0.091\,$\pm$\,0.007			& 1.003\,$\pm$\,0.000	& 0.964\,$\pm$\,0.003	\\
\phantom{1}9.050	& $1/2^+$	& 0.000	& $1/2^-$	& 						& 1.000			& 1.000	\\
\phantom{1}9.152	& $3/2^-$	& 0.000	& $1/2^-$	& -0.015\,$^+_-\,^{0.019}_{0.041}$	& 1.012\,$\pm$\,0.003	& 0.862\,$^+_-\,^{0.029}_{0.038}$	\\
\phantom{1}9.760	& $5/2^-$	& 0.000	& $1/2^-$	& 						& 0.858			& 0.999	\\
\phantom{1}9.925	& $3/2^-$	& 0.000	& $1/2^-$	& 						& 1.011			& 0.875	\\
10.066		& $3/2^+$	& 0.000	& $1/2^-$	& 						& 1.011			& 0.875	\\
10.702		& $3/2^-$	& 0.000	& $1/2^-$	& -0.180\,$^+_-\,^{0.002}_{0.006}$	& 1.025\,$\pm$\,0.001	& 0.704\,$^+_-\,^{0.002}_{0.006}$	\\
10.804		& $3/2^+$	& 0.000	& $1/2^-$	& +0.02\,$\pm$\,0.01			& 1.009\,$\pm$\,0.001	& 0.892\,$\pm$\,0.001	\\
\end{tabular}
\end{ruledtabular}
\footnotetext[1]{Krane-Steffen phase convention}
\end{table*}
\begin{table*}[t]
\caption{\label{tab:angular_corr_11B} Angular correlations for $^{11}$B transitions used in the analysis. Level parameters from \Ref{Kelley12-NPA} and mixing ratios from \Ref{Rusev09-PRC}.} 
\begin{ruledtabular}
\begin{tabular}{*{7}{c}}
$E_x$ (MeV)	& $J^\pi_x$	& $E_f$ (MeV)		& $J^\pi_f$	&Mixing ratio\footnotemark[1]& $W(127^\circ)$	& $W(90^\circ)$		\\
\colrule
4.445		& $5/2^-$	& 0.000	& $3/2^-$	& +0.158\,$^+_-\,^{0.025}_{0.021}$	& 1.000					& 0.998\,$^+_-\,^{0.002}_{0.004}$	\\
5.020		& $3/2^-$	& 0.000	& $3/2^-$	& -0.036\,$\pm$\,0.013			& 1.005\,$\pm$\,0.001			& 0.941\,$\pm$\,0.007	\\
5.020		& $3/2^-$	& 2.125	& $1/2^-$	& -0.19\,$^+_-\,^{0.10}_{0.17}$	& 0.989\,$^+_-\,^{0.002}_{0.003}$	& 1.134\,$^+_-\,^{0.036}_{0.030}$	\\
7.286		& $5/2^+$	& 0.000	& $3/2^-$	& +0.001\,$^+_-\,^{0.022}_{0.021}$	& 1.006\,$\pm$\,0.001			& 0.931\,$^+_-\,^{0.015}_{0.016}$	\\
7.286		& $5/2^+$	& 2.125	& $1/2^-$	& +0.028\,$^+_-\,^{0.073}_{0.075}$	& 1.005\,$^+_-\,^{0.003}_{0.002}$	& 0.940\,$^+_-\,^{0.031}_{0.035}$	\\
7.978		& $3/2^+$	& 0.000	& $3/2^-$	& 						& 1.007					& 0.92	\\
7.978		& $3/2^+$	& 2.125	& $1/2^-$	& 						& 0.992					& 1.10	\\
8.920		& $5/2^-$	& 0.000	& $3/2^-$	& 0.000\,$\pm$\,0.014			& 1.006\,$\pm$\,0.001			& 0.93\,$\pm$\,0.01	\\
\end{tabular}
\end{ruledtabular}
\footnotetext[1]{Krane-Steffen phase convention}
\end{table*}
The formalism describing the angular distribution in NRF is equivalent  to the theory of $\gamma-\gamma$ angular correlations \cite{Siegbahn68-book}. $W(\theta)$ is the probability of emission of the deexcitation photon at an angle of $\theta$ with respect to the direction of the absorbed photon.
$W(\theta)$ can be calculated from the level spin and mixing ratios taken from the literature as e.g. from \Ref{Kneissl06-JPG}. The factors used in the analysis of the $^{15}$N transitions are based on the compilation \cite{AjzenbergSelove91-NPA}, and are shown in \tab{tab:angular_corr_15N}. The calculations include the solid angle of the detectors, and the uncertainties includes angular uncertainty and the uncertainty of the mixing ratios. On average in the case of the $127^\circ$ detectors this factor differs from unity by less than 1\%; in the case of the $90^\circ$ detectors the difference from unity varies up to 25\%.

Similar factors have been calculated based on level parameters from \Ref{Kelley12-NPA} and mixing ratios from \Ref{Rusev09-PRC} for the $^{11}$B transitions (\tab{tab:angular_corr_11B}) used in the bremsstrahlung flux determination.

\subsection{\label{sec:flux} Bremsstrahlung flux}

The bremsstrahlung was created by an electron beam impinging onto a 12.5\,$\mu$m thick niobium radiator foil. A 10\,cm thick aluminum absorber (beam hardener) was inserted in the path of the photons to reduce the number of low-energy $\gamma$ rays.

The flux of the impinging $\gamma$-rays at the energies of the $^{15}$N levels was determined relative to those at the energies of $^{11}$B levels.
A theory-based \cite{Schiff51-PR} interpolation curve was used between the energies of the $^{11}$B lines, and as an extrapolation to determine the impinging photon flux at higher energies. To calculate this curve, in addition to the atomic number of the radiator, only the energy of the impinging electrons, thus the end-point energy is needed \cite{Schiff51-PR}. For precise determination of the end-point energy of the bremsstrahlung, it was measured independently from the electron beam diagnostics. 

\begin{figure}[t]
\includegraphics[width=0.99\columnwidth]{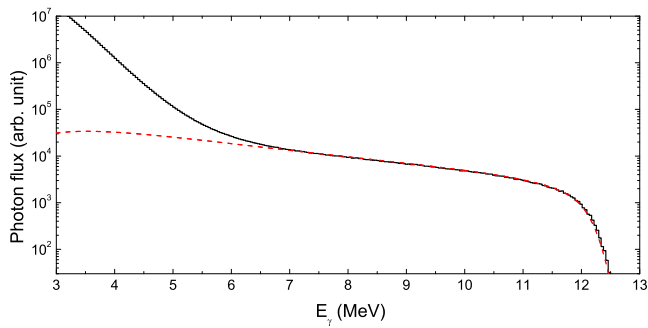}
\vspace{-5mm}
\caption{\label{fig:breakup_flux} (Color online) Comparison of the experimental (black histogram) and calculated (red dashed curve) bremsstrahlung energy distribution. For details see text.}
\vspace{-3mm}
\end{figure}

Right before the photon scattering site a deuteron target was placed in the path of the $\gamma$-rays. The spectrum of the emitted protons from this deuterated-polyethylene foil was measured by four ion implanted silicon detectors placed at 90$^\circ$ with respect to the beam axis. Using the known kinematics of the D($\gamma$,p)n reaction and the binding energy of the deuteron, the proton energy distribution was converted to $\gamma$-ray energy distribution (\fig{fig:breakup_flux}). This conversion took into account correction factors, i.\,e., the energy loss of the protons in the polyethylene, and the difference of the stopping power of protons and alpha particles used for the energy calibration of the detectors \cite{Erhard09-Diss}. The end-point energy from this measurement was used in the bremsstrahlung spectrum calculation.
In the analysis the calculated bremsstrahlung flux folded  by a hardener correction \cite{Mass11} was used to take into account the absorption in the beam hardener.

To independently validate the calculated brems\-strah\-lung spectrum, it was folded by the D($\gamma$,p)n reaction cross section \cite{ENDF-B-VII-1} and by the experimental resolution of the proton detection. The measured bremsstrahlung energy distribution derived from the proton distribution is well reproduced by the calculation (\fig{fig:breakup_flux}). The deviation at lower energies is caused by electrons, positrons, and $\gamma$ rays originating from the polyethylene foil reaching the Si detectors.

The $^{15}$N targets were always combined with a boron pill with known number of $^{11}$B nuclei. From the well-known scattering cross sections of levels in $^{11}$B \cite{Mohr02-AIP_conf} using the known $\gamma$ detection efficiency and angular correlation (\tab{tab:angular_corr_11B}) the flux of $\gamma$ rays was obtained at the energies of the boron levels. The photon flux at the $^{15}$N levels was determined by scaling the previously deduced photon energy distribution to the boron values in each run.

\begin{figure}[t]
\includegraphics[width=0.99\columnwidth]{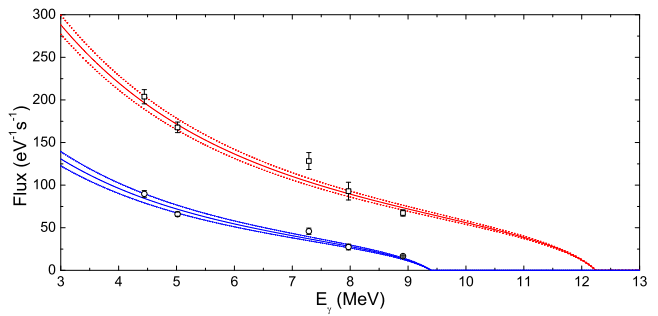}
\vspace{-6mm}
\caption{\label{fig:abs_flux} (Color online) Average bremsstrahlung flux experienced by target~\#4 during the different runs. The previously calculated distributions matching the experimental spectrum from the D($\gamma$,p)n reaction are scaled to the measured $^{11}$B values, and are potted by red (blue) lines for end-point of 12.6\,MeV (9.8\,MeV) with a 1$\sigma$ uncertainty bands, respectively.}
\end{figure}

Four $^{11}$B levels were used to fit the photon flux. In case of three levels out of these four not only the ground state transition, but also the transition to the first excited state was observed. The impinging flux was determined both from the elastic and inelastic photon scattering and consistent results were obtained both from transitions to the ground state and to the first excited state. This confirms independently the correctness of the efficiency calibration. The flux at a given level was then obtained by weighted averaging. In the averaging the statistical and systematic uncertainties were treated separately. The statistical uncertainty was used in the weighting, and the systematic uncertainty was quadratically added afterward.
The obtained flux values is plotted in \fig{fig:abs_flux} together with the fits.
The value deduced for the 7.280-MeV level overestimates the photon flux in all runs. This behavior was also observed in previous measurements \cite{Schwengner07-PRC, Makinaga10-PRC, Massarczyk13-PRC}. Therefore, this value was left out from the fit. This resulted in a minor reduction of the flux (about 1\%), because the level width of the 7.280-MeV level has a relatively large uncertainty and thus little weight in the fit compared to the other data points.

Because the flux is determined relative to the $^{11}$B values from the same experimental $\gamma$ spectrum, a precise knowledge of the dead time of the measurement is not necessary. In the present measurement the dead time of the counting setup was found to be negligible, but even if this were not the case, it would affect both the $^{11}$B peaks and the peaks of interest in the spectra, thus eliminating the effect.

\subsection{\label{sec:target} Targets}

Targets were produced from two different solid nitrogen compounds [ammonium nitrate (NH$_4$NO$_3$) and ammonium chloride (NH$_4$Cl)] enriched in $^{15}$N. In total, four targets were produced, one thicker and one thinner from both materials, to investigate and exclude self-absorption effects. The target powders were compressed in a cylinder to form disks with 2-cm diameter and enclosed vacuum tightly between two thin polyethylene films forming small bags. Small pinholes have been made at one of the corners of the bags to circumvent their explosion in the vacuum chamber. The pills made of compressed nitrogen compounds were mechanically stable. 

The masses of the produced pills were measured with an uncertainty of 0.2\,mg, which is 0.04\% precision for the pill with the smallest mass. 
The material certificate gives no error to the enrichment value but quotes it on a tenths of a percent level, therefore we assigned an uncertainty of 0.1\% to these values.
From both materials the nominal 2\,g were ordered. The measured weight of the slightly hygroscopic NH$_4$NO$_3$ was 7\% more than the nominal value, while the NH$_4$Cl was within 0.5\% of the ordered amount. 
Assuming the extra mass to be water in the NH$_4$NO$_3$, the number of $^{16}$O nuclei have been determined. From the two well-known level widths of $^{16}$O \cite{Tilley93-NPA}, the assumption on the water content was cross checked. These $^{16}$O level widths are well reproduced within 5\% experimental uncertainty, considering the above mentioned water content. The NH$_4$NO$_3$ material captures moisture from the atmosphere only above its critical relative humidity of 59.4\%. No mass change was observed during the pill production, and after their enclosure into the plastic bags. Before and between the irradiation the pills were stored in a vacuum desiccator, and their mass was regularly measured. There was no observable mass change within the 0.2-mg precision of the scale. Therefore, we assume that no water escaped from the target into the vacuum system.
\begin{table*}[t]
\caption{\label{tab:targets} Properties of the targets used}
\begin{ruledtabular}
\begin{tabular}{l*{5}{c}}
\begin{tabular}{l} Target \\ number \end{tabular}&
Material&
\begin{tabular}{c} Enrichment\footnotemark[1] \\ in $^{15}$N (\%) \end{tabular} &
Mass (mg)&
\begin{tabular}{c} Areal density of \\ $^{15}$N (10$^{21}$ cm$^{-2}$) \end{tabular}&
\begin{tabular}{c} Areal density of  \\ $^{16}$O (10$^{21}$ cm$^{-2}$) \end{tabular}\\
\colrule
\#1	& NH$_4$Cl				& 99.2\,$\pm$\,0.1	& \phantom{1}507.4\,$\pm$\,0.2	& 2.50\,$\pm$\,0.01	&	\\
\#2	& NH$_4$Cl				& 99.2\,$\pm$\,0.1	& 1503.3\,$\pm$\,0.2	& 7.42\,$\pm$\,0.04	&	\\
\#3	& NH$_4$NO$_3$ + H$_2$O	& 98.2\,$\pm$\,0.1	& \phantom{1}514.6\,$\pm$\,0.2	& 3.13\,$\pm$\,0.16	& 5.3\,$\pm$\,0.3\\
\#4	& NH$_4$NO$_3$ + H$_2$O	& 98.2\,$\pm$\,0.1	& 1624.3\,$\pm$\,0.2	& 9.9\,$\pm$\,0.5	& 16.6\,$\pm$\,0.8\\
\end{tabular}
\end{ruledtabular}
\footnotetext[1]{From the material certificate}
\end{table*}

As a conservative estimate in the analysis an uncertainty of 0.5\% was adopted for the number of target nuclei in case of the NH$_4$Cl pills, and 5\% in case of the NH$_4$NO$_3$ pills. The later bigger uncertainty reflects the unknown stoichiometry, and the level of uncertainty on which the $^{16}$O level widths used in the water content estimation were determined.
Detailed target properties are listed in \tab{tab:targets}.

For the boron pill production amorphous metallic boron powder enriched in $^{11}$B was used (99.5\% enrichment). The row material (200\,mg) together with small amount of polyvinyl-alcohol (dissolved in 10\,mg water) was compressed in the same tool as the nitrogen pills. After the pill dried the polyvinyl-alcohol stabilized it. Finally the $^{11}$B pill was also enclosed between thin polyethylene films. The weight of the pill+plastic is regularly checked, and found to be stable.
No lines from $^{16}$O were observed during the irradiation of the NH$_4$Cl pill combined with the boron pill, therefore we can assume that no water left from the solution in the $^{11}$B pill. 

The areal density of $^{15}$N and $^{11}$B nuclei used in the analysis is determined from the mass, molar mass, enrichment of the material and the area of the pills.
The beam spot is larger than the pills and homogeneously covers the targets, therefore the uncertainty of the geometrical size of the pills does not influence the photon flux determination.
\begin{figure*}[t]
\includegraphics[width=0.99\textwidth]{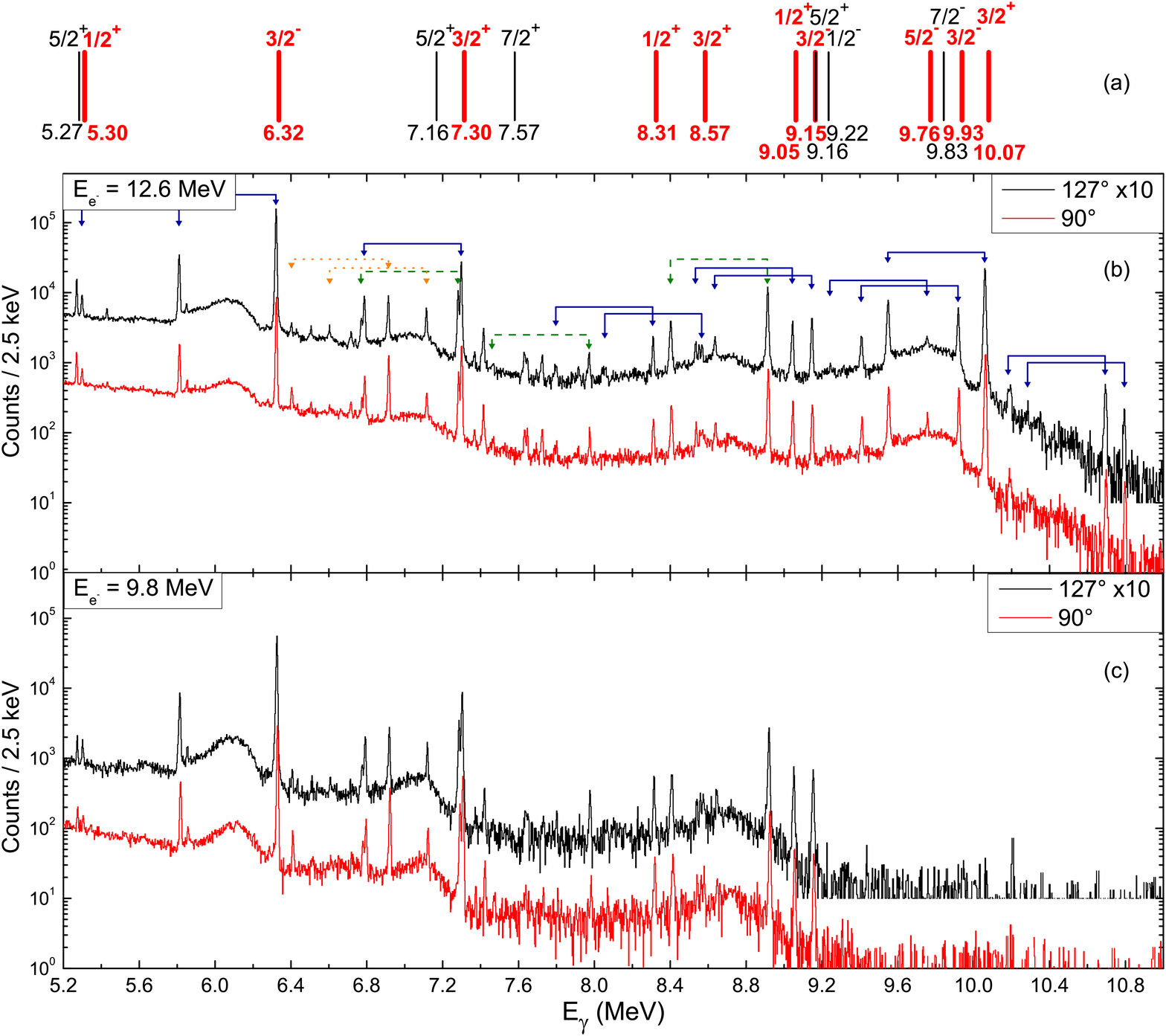}
\caption{\label{fig:spectrum} (Color online) (a): Level scheme of $^{15}$N from the first excited state up to the proton separation threshold at 10.21\,MeV. The spin and parities of the levels, and the level energies in MeV are taken from \cite{AjzenbergSelove91-NPA}. The investigated levels are marked by bold lines in red.
(b): Spectra recorded by the HPGe detectors placed at different angles, during the irradiation of target~\#4 with bremsstrahlung created by electrons of 12.6-MeV kinetic energy. (c): Same as (b), but with bremsstrahlung endpoint energy of 9.8\,MeV. Peaks correspond to $^{15}$N (blue solid),  $^{11}$B (green dashed), and  $^{16}$O (orange dotted) transitions, and the respective single escape peaks are marked. A few weak double escape peaks are also observed, but not marked in the spectra. The spectra recorded at 127$^\circ$ were multiplied by a factor of 10 for clearer view.}
\end{figure*}

\subsection{\label{sec:feeding} Feeding}

An experimental difficulty may arise in the data analysis of a typical NRF experiment. Namely, the measured quantity $I_{\sigma+f}$ derived from the peak area does not represent purely the energy and angle integrated resonant scattering cross section ($I_{\sigma}$), but it may contain a feeding part ($I_f$) from higher lying states, in addition to $I_\sigma$. $I_f$ is proportional to the population of the feeding level. Because the population of a level is proportional to its scattering cross section, the feeding of level $x$ from a higher lying level $y$ can be estimated as
\begin{equation}
I_f^x = \frac{b_x^y}{b_0^y}  \frac{\Phi(E_y)}{\Phi(E_x)} I_{\sigma}^y ,
 \label{eq:feeding}
\end{equation}
where $b_x^y$ and $b_0^y$ are the gamma branching ratios of level $y$ to level $x$ and to the ground state, respectively. For the estimation the branching rations have been adopted from \cite{AjzenbergSelove91-NPA}. While $\Phi(E_y)$ and $\Phi(E_x)$ are the measured photon flux at the energy of the levels $y$ and $x$, respectively.

\section{\label{sec:measurement} Measurements}

Spectra were recorded with each target with bremsstrahlung produced by electrons with kinetic energy of 12.6\,MeV. Another experiment was carried out with a bremsstrahlung end-point energy of 9.8\,MeV to investigate and circumvent the feeding.
Spectra recorded at two different angles with the two different bremsstrahlung end-point energies are shown in \fig{fig:spectrum}, where also the $^{15}$N levels are shown from the first excited state up to the proton separation energy, 10.207\,MeV \cite{AjzenbergSelove91-NPA}. 
Levels shown in black have a low ground-state branching, and hence low scattering cross sections such that this method and experimental setup is not sensitive enough to study them.
The levels shown in bold red are investigated in this work; gamma peaks in the spectra correspond to all highlighted $^{15}$N levels.

The peaks were fitted by Gaussian functions plus linear background.
In each run the efficiency and angular distribution corrected peak area, thus the number of excited $^{15}$N and $^{11}$B atoms have been determined. Consistent results were obtained for both angles. The weighted average of these values were used in the analysis.

The self-absorption is proportional to the scattering cross section and target mass, therefore it would be observable in the yield of the 6.324-MeV peak.
The target mass and photon flux corrected intensity of the peaks were within their statistical uncertainty regardless of the mass of the pills or target composition. Therefore, detectable self-absorption effects were not present, and self-absorption correction was not applied. 
Also consistent $I_{\sigma+f}$ values for each level prior the feeding correction were obtained from each run with bremsstrahlung end-point energy of 12.6\,MeV. The weighted averages of the obtained values are presented in column five of \tab{tab:XS}.

Irradiation with bremsstrahlung end-point energy of 9.8\,MeV have only been done for target~\#4; the obtained $I_{\sigma+f}$ values are presented in the fourth column of \tab{tab:XS}.
\begin{table*}[t]
\caption{\label{tab:XS} Measured scattering cross sections without and with feeding correction from irradiations with different bremsstrahlung end-point energies. Level energies, spins, parities and branching ratios are taken from \cite{AjzenbergSelove91-NPA}. The spin and parity of the ground state of $^{15}$N is $1/2^-$.}
\begin{ruledtabular}
\begin{tabular}{*{7}{c}}
\multirow{2}{*}{$E_x$ (MeV)}	& \multirow{2}{*}{$J^\pi_x$}	& \multirow{2}{*}{$b$ (\%)}	& $I_{\sigma+f}$ (eV\,b)		& $I_{\sigma+f}$ (eV\,b)		& $I_\sigma$ (eV\,b)			& $I_\sigma$ (eV\,b)			\\
					& 					& 					& $E_\mathrm{e^-} = 9.8$\,MeV&$E_\mathrm{e^-} = 12.6$\,MeV& $E_\mathrm{e^-} = 9.8$\,MeV	&$E_\mathrm{e^-} = 12.6$\,MeV 	\\
\colrule
\phantom{1}5.298			& $1/2^+$				& 100					& 6.9\,$\pm$\,0.8			& 13.1\,$\pm$\,0.8 		& 6.0\, $^+_-$\,$^{0.8}_{2.3}$	& 						\\
\phantom{1}6.324			& $3/2^-$				& 100					& 558\,$\pm$\,46			& 560\,$\pm$\,18 			& 557\,$\pm$\,46 				& 554\,$^+_-$\,$^{18}_{22}$ 		\\
\phantom{1}7.301			& $3/2^+$				& 99.3\,$\pm$\,0.7			& 160\,$\pm$\,13			& 159\,$\pm$\,5 			& 159\,$\pm$\,13 				& 156\,$^+_-$\,$^{5}_{10}$		\\
\phantom{1}8.313			& $1/2^+$				& 79\,$\pm$\,2			& 15.8\,$\pm$\,1.6			& 16.8\,$\pm$\,1.1			& 15.8\,$^+_-$\,$^{1.6}_{1.8}$	& 16.4\,$^+_-$\,$^{1.1}_{7.9}$	\\
\phantom{1}8.571 			& $3/2^+$ 				& 33\,$\pm$\,2			& 10.8\,$\pm$\,1.6 		& 8.7\,$\pm$\,0.8 			& 10.8\,$\pm$\,1.6 			& 						\\ 
					&					&\footnotemark[1]65\,$\pm$\,3	&\footnotemark[1]22\,$\pm$\,4 	&\footnotemark[1]18.3\,$\pm$\,1.5&\footnotemark[1]22\,$\pm$\,4	&\footnotemark[1]18.3\,$^+_-$\,$^{\phantom{1}1.5}_{11.6}$ \\ 
\phantom{1}9.050			& $1/2^+$				& 92\,$\pm$\,3			& 					& 39.8\,$\pm$\,1.8			& 						& 39.7\,$\pm$\,1.8				\\
\phantom{1}9.152			& $3/2^-$				& 100\,$\pm$\,3			& 					& 52.5\,$\pm$\,2.4			& 						& 52.4\,$\pm$\,2.4				\\
\phantom{1}9.760			& $5/2^-$				& 81.5\,$\pm$\,2.8			&					& 15.7\,$\pm$\,1.5			& 						& 15.7\,$\pm$\,1.5				\\
\phantom{1}9.925			& $3/2^-$				& 77.6\,$\pm$\,1.9			& 					& 103\,$\pm$\,4			& 						& 103\,$\pm$\,4				\\
10.066				& $3/2^+$				& 96.0\,$\pm$\,0.7			& 					& 426\,$\pm$\,14			& 						& 426\,$\pm$\,14				\\
10.702				& $3/2^-$				& 52.6\,$\pm$\,0.8			& 					& 14.5\,$\pm$\,1.2			& 						& 14.5\,$\pm$\,1.2				\\
10.804				& $3/2^+$				& 51.5\,$\pm$\,0.4			& 					& 6.8\,$\pm$\,0.8			& 						& 6.8\,$\pm$\,0.8 				\\
\end{tabular}
\end{ruledtabular}
\footnotetext[1]{Transition to the $5/2^+$ first excited state.}
\end{table*}

In column six and seven of \tab{tab:XS} the obtained integrated resonant scattering cross sections corrected for feeding are presented.
Where only upper limits were available for the gamma branchings \cite{AjzenbergSelove91-NPA}, only the quoted lower error bar of the value is affected by those resulting in asymmetric uncertainties. The feeding transitions are not observed in our spectra, because of the high background in the low-energy region where they are expected to appear. For the intensity of branching transitions upper limits can be calculated from the recorded spectra, but those are much higher than the literature upper limits. In the feeding estimate, the literature gamma branching ratios from \cite{AjzenbergSelove91-NPA} were used.

The first excited state (5.270\,MeV, $5/2^+$) is mainly populated through feeding in cases of both bremsstrahlung end-point energies. Even if the corresponding transitions were visible in the spectra, scattering cross section and level width information can therefore not be extracted.

In the case of the irradiation with the higher brems\-strahlung end-point energy sizable feeding contributions steaming from the 10.066-MeV level to the lower lying states are expected. In case of the 5.298-MeV and 8.571-MeV levels, the estimated feeding contribution dominates the peak area, thus no scattering cross sections were derived for these levels from the irradiations with bremsstrahlung end-point energy of 12.6\,MeV.

This problem is circumvented by the experiment with bremsstrahlung end-point energy of 9.8\,MeV. In this later measurement the 10.066-MeV level was not excited, and the feeding of the investigated levels was drastically reduced.

Although, in the spectra recorded during the irradiation with the lower bremsstrahlung end-point energy, peaks corresponding to the 9.050-MeV and 9.152-MeV levels are also visible, no scattering cross section was derived. These levels were excited by the falling edge of the bremsstrahlung distribution, resulting in an uncertain photon flux.

After subtraction of the feeding contribution [\eq{eq:feeding}], consistent scattering cross sections from the irradiations by different bremsstrahlung end-point energies are obtained (\tab{tab:XS}). 
Averaging of these values does not give higher precision, because the systematical error dominates the final uncertainty of the derived scattering cross sections, and those are common for both irradiations. $\Gamma_0$ values were calculated using the scattering cross section with higher precision.

\section{\label{sec:results} Results and discussion}

\subsection{Levels below the proton separation threshold}

The present partial gamma widths to the ground state ($\Gamma_0$) are deduced from the integrated resonant scattering cross sections using \eq{eq:I_elastic}, and compared to the literature in \tab{tab:G0}.

\begin{description}[leftmargin=0em]
\item[5.298\,MeV] 
The level lifetime of this level is given in the literature, 25\,$\pm$\,7\,fs \cite{AjzenbergSelove91-NPA}. The given literature level width in \tab{tab:G0} was calculated from this value. The present gamma width is consistent with similar precision, although its lower error bar is increased from the feeding.
\item[6.324\,MeV] 
The present level width is consistent with the literature value, but more precise. However, in the compilation \cite{AjzenbergSelove91-NPA} only the most precise value from \cite{Moreh81-PRC} is presented. There are also few other values from the same authors. {\it I.\,e.} $\Gamma_0 = 3.1 \pm 0.3$\,eV is quoted in \cite{Moreh75-NPA}, later with a new measurement this value was revised and $\Gamma_0 = 2.9 \pm 0.3$\,eV was quoted in \cite{Moreh76-NPA} considered to be more accurate.
\item[7.301\,MeV]
This level is the isospin mirror of the $E_x = 6.792$\,MeV, $3/2^+$ level in $^{15}$O. Although the present value has an asymmetric error bar from the feeding, it is consistent with the literature value and has higher precision. 
\item[8.313\,MeV] 
The gamma width of this level was calculated from the irradiation by the lower bremsstrahlung end-point energy, because the 10.066-MeV level may have sizable feeding. The obtained level width is consistent with the previous one \cite{Moreh81-PRC} but has a greatly improved precision.
\item[8.571\,MeV] 
The gamma width of this level was also calculated from the irradiation by the lower bremsstrahlung end-point energy. Not only the ground state transition but the transition to the first excited state was also observed. The presented $\Gamma_0$ is the weighted average of the consistent values obtained from the elastic and inelastic scattering cross sections, because both of them are related to $\Gamma_0$ by \eq{eq:I_from_width}. The precision of $\Gamma_0$ is greatly improved compared to the literature value.
\item[9.050\,MeV, 9.152\,MeV, 9.760\,MeV, 9.925\,MeV]
The gamma widths of these levels were deduced from the irradiation by the higher bremsstrahlung end-point energy. Consistent values with the literature are obtained, but with higher precision.
\item[10.066\,MeV] 
The obtained present $\Gamma_0$ value and that in the compilation \cite{AjzenbergSelove91-NPA} are consistent with each other, however, the present value is more precise. The value in the compilation (see. \tab{tab:G0}) rely on only one experimental data set from \cite{Moreh81-PRC}. However, the ground-state level width of the 10.066-MeV level was reported to be $\Gamma_0 = 4.2 \pm 1.5$\,eV in \cite{Patrick76-JPG}. This latter value is not used in the compilation, possibly because of the much lower precision.
\end{description}

\begin{table}[h]
\caption{\label{tab:G0} Measured partial gamma widths to the ground state of levels below the proton separation threshold.}
\begin{ruledtabular}
\begin{tabular}{*{4}{c}}
\multirow{2}{*}{$E_x$ (MeV)}	& $\Gamma_0$ (eV)			& $\Gamma_0$ (eV)				& 					\\
					& This work					& Literature						& Ref.					\\
\colrule
\phantom{1}5.298			& 0.044\,$^+_-$\,$^{0.006}_{0.017}$&0.026\,$^+_-$\,$^{0.010}_{0.006}$	& \cite{AjzenbergSelove91-NPA}\footnotemark[1]	\\
\phantom{1}6.324			& 2.88\,$^+_-$\,$^{0.09}_{0.11}$	& 3.12\,$\pm$\,0.18\footnotemark[2] 		& \cite{Moreh81-PRC}		\\
\phantom{1}7.301			& 1.09\,$^+_-$\,$^{0.04}_{0.07}$	& 1.08\,$\pm$\,0.08				& \cite{Moreh81-PRC}		\\
\phantom{1}8.313			& 0.36\,$\pm$\,0.04			& 0.3\,$\pm$\,0.2					& \cite{Moreh81-PRC}		\\
\phantom{1}8.571			& 0.32\,$\pm$\,0.04			& 0.3\,$\pm$\,0.3					& \cite{Moreh81-PRC}		\\
\phantom{1}9.050			& 0.92\,$\pm$\,0.05			& 1.2\,$\pm$\,0.2					& \cite{Moreh81-PRC}		\\
\phantom{1}9.152			& 0.57\,$\pm$\,0.03			& 0.47\,$\pm$\,0.12				& \cite{Moreh81-PRC}		\\
\phantom{1}9.760			& 0.160\,$\pm$\,0.016			& 0.21\,$\pm$\,0.07				& \cite{Moreh81-PRC}		\\
\phantom{1}9.925			& 1.70\,$\pm$\,0.08			& 1.6\,$\pm$\,0.2					& \cite{Moreh81-PRC}		\\
10.066				& 5.85\,$\pm$\,0.20			& 6.3\,$\pm$\,0.4\footnotemark[3]		& \cite{Moreh81-PRC}		\\
\end{tabular}
\end{ruledtabular}
\footnotetext[1]{Level lifetime of 25\,$\pm$\,7\,fs is given.}
\footnotetext[2]{$\Gamma_0 = 3.1 \pm 0.3$\,eV is quoted in \cite{Moreh75-NPA}, and $\Gamma_0 = 2.9 \pm 0.3$\,eV is quoted in \cite{Moreh76-NPA}.}
\footnotetext[3]{Others quote $\Gamma_0 = 4.2 \pm 1.5$\,eV, deduced from $I_\sigma = 320 \pm 100$\,eV\,barn \cite{Patrick76-JPG}.}
\end{table}

\subsection{Levels above the proton separation threshold}

\begin{table}[t]
\caption{\label{tab:res2} Measured total gamma widths of levels above the proton separation threshold. Evaluated total gamma and proton widths from \cite{AjzenbergSelove91-NPA}.}
\begin{ruledtabular}
\begin{tabular}{*{4}{c}}
\multirow{2}{*}{$E_x$ (MeV)}	& $\Gamma_\gamma$ (eV)& $\Gamma_\gamma$ (eV)	& $\Gamma_p$ (eV)	\\
					& This work			& \Ref{AjzenbergSelove91-NPA}	& \Ref{AjzenbergSelove91-NPA}	\\
\colrule
10.702				& 0.78\,$\pm$\,0.06	& 0.37\,$\pm$\,0.07		& 200	\\
10.804				& 0.39\,$\pm$\,0.04	& 0.27\,$\pm$\,0.14		& 0.22\,$\pm$\,0.10	\\
\end{tabular}
\end{ruledtabular}
\end{table}

Elastic gamma scattering cross sections of two levels above the proton separation threshold (10.207\,MeV) of $^{15}$N were measured, too. The gamma widths of these levels are derived for the first time from a gamma scattering measurement without the need of the proton width of the given levels (\tab{tab:res2}).

In former works the resonance strengths in the $^{14}$C(p,$\gamma$)$^{15}$N reaction populating these levels were measured. The resonance strength ($\omega\gamma$) is related to the total gamma width ($\Gamma_\gamma$) in the case of these levels as
\begin{equation}
\omega\gamma = \frac{2J_x+1}{(2j_t+1)(2j_p+1)} \frac{\Gamma_p\Gamma_\gamma}{\Gamma_p+\Gamma_\gamma} ,
 \label{eq:omegagamma}
\end{equation}
where $J_x$ is the spin of the excited state in $^{15}$N, $j_t$, and $j_p$ are the spins of the target ($^{14}$C) and projectile, respectively, and  $\Gamma_p$ is the proton width of the given level. In this study $j_t = 0$ and $j_t = 1/2$, because the ground state of $^{14}$C is $0^+$ and the projectile is a proton.

The use of \eq{eq:omegagamma} for gamma width determination may require the  knowledge of the proton width.
\begin{enumerate}
\item If $\Gamma_p \gg \Gamma_\gamma$, then $\omega\gamma$ became proportional to the total gamma width, and $\Gamma_p$ can be neglected.
\item The absolute value of $\Gamma_p$ has to be known, when the two widths are on a comparable level.
\end{enumerate}
In case of the two investigated levels in this work, both of these cases appear. 

\begin{description}[leftmargin=0em]
\item[10.702\,MeV] 
This level fulfills the $\Gamma_p \gg \Gamma_\gamma$ criterion. Therefore, \eq{eq:omegagamma} becomes 
\begin{equation} 
\omega\gamma = 2 \Gamma_\gamma .
\end{equation} 
$\omega\gamma$ can be derived from the present measurement, and compared to the literature (\tab{tab:omegagamma}).
The measured new resonance strength is in agreement with \cite{Siefken69-NPA}, but in disagreement with the values reported in \cite{Beukens76-thesis,Gorres90-NPA}
\end{description}
\begin{table}[h]
\caption{\label{tab:omegagamma} Strength of the 527\,keV proton resonance in $^{14}$C(p,$\gamma$)$^{15}$N reaction ($E_x = 10.702$\,MeV).}
\begin{ruledtabular}
\begin{tabular}{*{4}{c}}
$\omega\gamma$  (eV)	& $\omega\gamma$  (eV)	& $\omega\gamma$  (eV)	& $\omega\gamma$  (eV)	\\
\Ref{Siefken69-NPA}	& \Ref{Beukens76-thesis}	& \Ref{Gorres90-NPA}	& This work	\\
\colrule
1.78\,$\pm$\,0.30		& 0.74\,$\pm$\,0.14 	& 0.84\,$\pm$\,0.13	& 1.55\,$\pm$\,0.13 \\
\end{tabular}
\end{ruledtabular}
\end{table}

\begin{description}[leftmargin=0em]
\item[10.804\,MeV] 
In the case of this level $\Gamma_p$ and $\Gamma_\gamma$ are comparable \cite{AjzenbergSelove91-NPA}. Even if the reported resonance strength has a precision of 25\% as given in the literature \cite{AjzenbergSelove91-NPA}, the huge uncertainty of $\Gamma_p$ \cite{AjzenbergSelove91-NPA} resulting in about a 50\% uncertainty of $\Gamma_\gamma$.
Our approach does not need the use of $\Gamma_p$. Therefore, we can quote a more accurate value, with about a 10\% uncertainty (see \tab{tab:res2}) that comes from the counting statistics and the uncertainty of the branching ratio.
\end{description}
Furthermore, we can turn around the argument in \eq{eq:omegagamma}, and use it to determine $\Gamma_p$ from the known $\omega\gamma$ values and from the new independent $\Gamma_\gamma$.
In this calculation the weighted average of the resonance strengths found in the literature was used (see \tab{tab:proton}).
\begin{table}[h]
\caption{\label{tab:proton} Strength of the 634\,keV proton resonance in $^{14}$C(p,$\gamma$)$^{15}$N reaction ($E_x = 10.804$\,MeV) from the literature used for the proton width calculation.}
\begin{ruledtabular}
\begin{tabular}{*{3}{c}|c}
$\omega\gamma$  (eV)	& $\omega\gamma$  (eV)	& $\omega\gamma$  (eV)	& $\omega\gamma$  (eV)	\\
\Ref{Siefken69-NPA}	& \Ref{Beukens76-thesis}	& \Ref{Gorres90-NPA}	& Average	\\
\colrule
0.23\,$\pm$\,0.04		& 0.24\,$\pm$\,0.06 	& 0.27\,$\pm$\,0.04	& 0.25\,$\pm$\,0.03 \\
\end{tabular}
\end{ruledtabular}
\end{table}
The resulting proton width in this work:
\begin{equation}
\Gamma_p(\mathrm{10.804\,MeV}) = 0.18 \pm 0.03 \mathrm{\,eV}
\end{equation}
 is consistent with the literature value \cite{AjzenbergSelove91-NPA} (\tab{tab:res2}), but more precise.

\section{\label{sec:sum} Summary}

Ground state gamma widths in $^{15}$N have been determined with NRF technique. Bremsstrahlung was used to excite the nuclear levels, and HPGe detectors with BGO shields to detect the scattered photons.
Solid nitrogen compounds enriched in $^{15}$N have been used, and consistent results are obtained for several values of target thickness and composition, and several bremsstrahlung end-point energies.

The results are consistent with the literature values, but with much improved precision. The new experimental data can be used as a reference for future investigations.

The resonance strength in the $^{14}$C(p,$\gamma$)$^{15}$N reaction populating the 10.702-MeV level in $^{15}$N was determined for the first time from a gamma scattering experiment, and found to be in agreement with a previous value \cite{Siefken69-NPA}, but in contradiction with others \cite{Beukens76-thesis, Gorres90-NPA}.

The proton width of the 10.804-MeV level was determined using the gamma width determined here and the resonance strengths from the literature. The obtained value has a higher precision than the previous one.

\begin{acknowledgments}
This work was supported by the Helmholtz Association through the Nuclear Astrophysics Virtual Institute (NAVI, HGF VH-VI-417), BMBF (NupNET NEDENSAA 05P12 ODNUG), DFG (BE4100/2-1), and OTKA (K101328, K108459).
\end{acknowledgments}


\end{document}